\documentclass[useAMS,usenatbib]{mn2e}
\input{epsf}

\title[Variable stars in the field of the old open cluster Melotte 66]{Variable stars in the field of the old open cluster Melotte~66}

\author[K. Zloczewski, J. Kaluzny, W. Krzeminski, A. Olech and I. B. Thompson]
{K. Zloczewski$^1$\thanks{E-mail: kzlocz@camk.edu.pl (KZ); jka@camk.edu.pl (JK);\newline wojtek@lco.cl (WK); olech@camk.edu.pl (AO); ian@ociw.edu (IBT)}, J. Kaluzny$^1$, W. Krzeminski$^2$, A. Olech$^1$ and I. B. Thompson$^3$
\\
$^1$Nicolaus Copernicus Astronomical Center, ul. Bartycka 18, 00-716 Warsaw, Poland\\
$^2$Las Campanas Observatory, Carnegie Observatories, Casilla 601, La Serena, Chile\\
$^3$The Observatories of the Carnegie Institution of Washington, 813 Santa Barbara St. Pasadena, CA 91101
}

\begin{document}

\date{Accepted --. Received --; in original form 2007 April 23}

\pagerange{\pageref{firstpage}--\pageref{lastpage}} \pubyear{2007}

\maketitle

\label{firstpage}

\begin{abstract}

We report the  results of photometric monitoring of the Melotte 66 field
in {\it BVI} filters. Ten variables were identified with nine being new
discoveries. The sample includes eight eclipsing binaries of which four
are W UMa type stars, one star is a candidate blue straggler. All four
contact binaries are likely members of the cluster based on their
estimated distances. Ten blue stars with $U-B<-0.3$ were detected inside
a $14.8 \times 22.8$ arcmin$^{2}$ field centred on the cluster. Time
series photometry for 7 of them showed no evidence for any  variability.
The brightest object in the sample of blue stars is a promising
candidate for a hot subdwarf belonging to the cluster. We show that the
anomalously wide main sequence of the cluster, reported in some earlier
studies, results from a combination of two effects: variable reddening
occuring across the cluster field  and the presence of a rich population
of binary stars in the cluster itself. The density profile of the
cluster field is derived and the total number of member stars with
$16<V<21$ or $2.8<M_{V}<7.8$ is estimated conservatively at about 1100.

\end{abstract}

\begin{keywords}
binaries: close -- open cluster and associations: individual: Melotte 66
\end{keywords}

\section{Introduction}

Melotte 66 ($\alpha = 07^{\rmn{h}} 26^{\rmn{m}} 21.^{\rmn{s}}9$, 
$\delta = -47\degr 41\arcmin 19\arcsec$, J2000) belongs to a small
sample of old galactic open clusters (Kassis et al. 1997). It is located
at a relatively large galactic latitude ($l=259\fdg6$, $b=-14\fdg3$). A
pioneering study, based on the photographic photometry was conducted by
Eggen \& Stoy (1962). Subsequent investigations based on photographic
and photoelectric photometry (Hawarden 1976;  Anthony-Twarog et al.
1979) were followed by several papers published in the last two decades
of the 20th century. Kassis et al. (1997) used deep $VI$ photometry to
derive an age of $4\pm 1$~Gyr and distance modulus $(m-M)_{\rmn{0}}=13.2
^{+0.3}_{-0.1}$. Anthony-Twarog et al. (1994) used $vbyH\beta$ data to
discuss the possible causes of an atypically wide cluster main sequence
which was originally noted by Anthony-Twarog et al. (1979). They
excluded  differential reddening across the cluster field as a cause of
this effect.

The cluster stands out from the sample of the known old open clusters in
its exceptionally low metallicity. Twarog et al. (1995)
derived $[\rmn{Fe/H}]=-0.53\pm 0.08$ based on $UBV$ photometry of
turnoff stars. Friel \& Janes (1993) obtained $[\rmn{Fe/H}]=-0.51\pm
0.11$ from a spectroscopic analysis of 4 giants while Gratton \&
Contarini (1994)  derived $[\rmn{Fe/H}]=-0.38\pm 0.15$ from high
resolution spectra of two giants.

This paper is a contribution to the systematic search for short period
variables in open clusters conducted by our group over the last two decades.
One of the goals is to establish a relation between age and a relative
frequency of occurrence of contact binaries in stellar clusters. A
summary of some of our earlier results can be found in  Rucinski (1998).
Here we report the results of a survey for variable stars in the field  of
Melotte 66 and also present deep CCD $UBVI$ photometry for the cluster.

\section[]{Observations and Reductions}

\begin{table*}
\begin{minipage}{120mm}
\caption{Summary of Melotte 66 observations}
\begin{tabular}{@{}l l c c c c c}
\hline
Run             & Dates         & Nights        & \multicolumn{3}{c}{ No of exposures}  & Median seeing \\
                &               & no            &   B   &  V    & I                     & in V filter   \\

\hline
FORD2-1         & 03-08.02.1992 & 5             & 8     & 59    & --                    & 1\arcsec5       \\
FORD2-2         & 20-24.03.1992 & 3             & --    & 12    & 44                    & 1\arcsec2       \\
TEK2-1          & 09-16.02.1992 & 7             & --    & 134   & 5                     & 1\arcsec5       \\
TEK2-2          & 13-18.03.1992 & 6             & 68    & 66    & --                    & 1\arcsec5       \\
\hline
\end{tabular}
\end{minipage}
\end{table*}

The field of Melotte 66 was surveyed for variable stars with the 1-m
Swope telescope at Las Campanas Observatory. The observations were
collected on a total of 21 nights during 4 observing runs during the
period February--March 1992. Two different cameras were used: a
$1024\times 1024$ Tektronics CCD with a scale of 0.61 arcsec
pixel$^{-1}$ (TEK2 camera with a $10.4\times 10.4$ arcmin$^{2}$ field)
and a $2048\times 2048$ Ford Aerospace CCD with a scale 0.435 arcsec
pixel$^{-1}$ (FORD2 camera with a $14.8\times 14.8$ arcmin$^{2}$ field).

Most of the images were collected with a $V$ filter with exposure times
ranging from 60 s to 480 s with a median value of 420 s. Exposures were
also obtained in the $B$ and $I$ bands. A summary log of the
observations is listed in Table 1. An additional set of $UBVI$
observations to calibrate the photometry and to construct a
color-magnitude diagram (CMD)  for the cluster field was obtained on the
nights of 1999 November 17--20 (UT) using the $2048\times 3150$ SITE3
camera. With a scale of 0.435 arcsec pixel$^{-1}$ this camera provides a
field of view of  $14.8\times 22.8$ arcmin$^{2}$. Several exposures of
different length were obtained in each of the four filters.  All images
were corrected for the known non-linearity of the SITE3 camera using the
procedure described in Hamuy et al. (2006).

\begin{figure*}
\begin{center}
   \leavevmode
   \epsfxsize=168mm
   \epsfbox[1498 477 4037 1482]{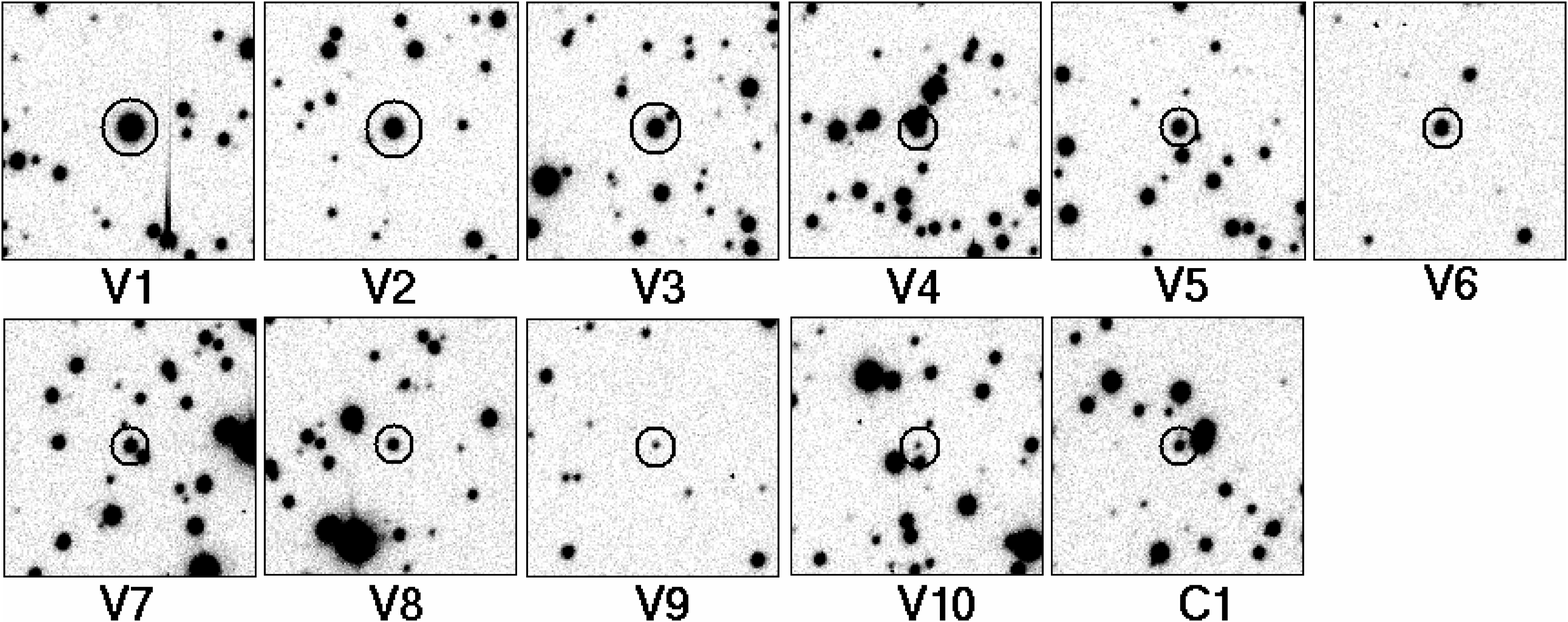}
\end{center}
\caption{Finding charts for variables V1--V10 and for the UV bright star C1.
Each chart is 1~arcmin on a side, with east to the left and north up.}
\label{charts}
\end{figure*}

Preliminary processing of the CCD frames was done with standard routines
in the {\sc IRAF-CCDPROC}\footnote{IRAF is distributed by the National
Optical Astronomy Observatory, which is operated by the Association of
Universities for Research in Astronomy, Inc., under a cooperative
agreement with the National Science Foundation.} package. Profile
photometry was extracted using the {\sc DAOPHOT/ALLSTAR} package
(Stetson 1987). For each camera/filter combination a "master" frame was
selected. This was used to create a  reference list of objects to be
measured in the fixed-position  mode on the remaining frames.
Instrumental magnitudes  were transformed to the system defined by the
master frame. The resulting data bases were searched for variable
objects using codes using the  analysis of variance (AoV) statistic 
(Schwarzenberg-Czerny 1996) and the {\sc AOVTRANS} algorithm
(Schwarzenberg-Czerny \& Beaulieu 2006). AoV periodograms were
calculated for periods spanning the range from 0.05 days to 20 days.
After the rejection of some spurious detections we ended up with a list
of 10 certain variable stars. Nine of these are new discoveries while
one is an eclipsing binary V345~Pup reported by Kaluzny and Shara
(1988). Equatorial coordinates of the variables were determined using
1651 stars from the USNO A-2 catalog (Monet et al. 1998) which were
identified on the  $V$-band image obtained with the  SITE3 camera. These
coordinates are listed in Table 2 along with the coordinates of a UV
bright star labelled C1 (see Sec. 3.1)  The last column of Table 2 gives
the angular distance from the cluster centre as determined from our data
 ($\rmn{RA}(2000)=07^{\rmn{h}} 26^{\rmn{m}} 21.^{\rmn{s}}9$ and
$\rmn{DEC}(2000)=-47\degr41\arcmin19\arcsec$). In Fig.~\ref{charts} we present
finding charts for the stars listed in Table 2.
~

\begin{table}
\caption{Equatorial coordinates of variables and a UV bright star
in the field of Melotte 66}
\begin{tabular}{c c c l}
\hline
ID      & $\alpha_{\rmn 2000}$ [deg] & $\delta_{\rmn 2000}$ [deg] & r[arcmin]\\
\hline
V1      & 111.59873             & $-$47.70060    &  0.78    \\
V2      & 111.53339             & $-$47.62291    &  4.58    \\
V3      & 111.61123             & $-$47.67409    &  1.18    \\
V4      & 111.57849             & $-$47.69664    &  0.71    \\
V5      & 111.56815             & $-$47.64329    &  2.87    \\
V6      & 111.72366             & $-$47.76692    &  7.12    \\
V7      & 111.56230             & $-$47.72299    &  2.38    \\
V8      & 111.54562             & $-$47.71756    &  2.54    \\
V9      & 111.71073             & $-$47.75736    &  6.35    \\
V10     & 111.63718             & $-$47.68840    &  1.85    \\
C1      & 111.57537             & $-$47.74433    &  3.31    \\
\hline
\end{tabular}
\end{table}

\subsection[]{Photometric calibration}

We have used the data collected on the night of 1999 November 18  to
calibrate our photometry. Observations of the cluster field were
bracketed by observations of three Landolt fields containing a total of
28 standard stars with  $BVI$ magnitudes (Landolt 1992; Stetson 2000) 
and 19 stars with  $U$ magnitudes (Landolt 1992). The T~Phe field was
observed twice while  each of the RU~149 and RU~152 fields   was
observed once. Standards were observed at air-masses spanning the range
1.12--1.24 while Melotte 66 was observed at an air-mass of about 1.1.
Average extinction coefficients for Las Campanas were assumed in
determining the linear transformations between the  instrumental and the
standard system. The total uncertainties of the zero points of our
photometry are about 0.02~mag for $BVI$ filters and about 0.1 mag for
the $U$ filter. These uncertainties include errors in the aperture
corrections  derived for the frames of Melotte 66. The relatively poor
quality of the $U$-band transformation can be explained by large
differences in the UV spectral responses of the SITe3 CCD camera and the
RCA~1P21 photomultiplier defining the $UBV$ system (Johnson 1963).
The median values of the formal internal errors of our photometry 
for the cluster range from $\sigma_{V}=0.010$ and $\sigma_{V-I}=0.021$
at $V=15.5$ to $\sigma_{V}=0.018$ and $\sigma_{V-I}=0.027$ at $V=19.5$.

\begin{figure}
\begin{center}
   \leavevmode
   \epsfxsize=84mm
   \epsfbox[53 53 355 385]{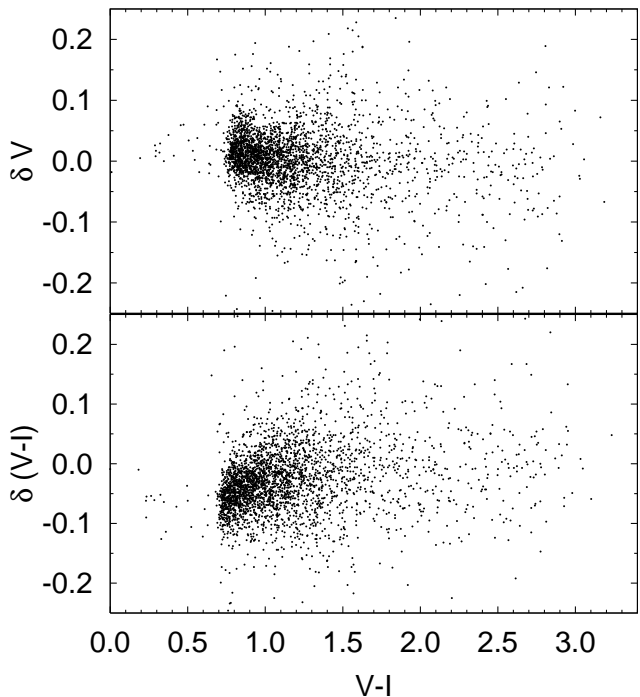}
\end{center}
\caption{Residuals of $V$ and $V-I$ for this work and Kassis et al. (1997) as a function of $V-I$.}
\label{comp}
\end{figure}

We compare our $VI$ photometry with the data taken from Kassis et al.
(1997) in Fig.~\ref{comp}. The mean difference (in the sense of "our"
measurements minus "theirs") is $-0.005\pm 0.001$ and $-0.023\pm 0.001$
for $V$ and $V-I$, respectively. There is no clear color dependence of
the residuals for the $V$ magnitudes. There are some systematic trends
for the $V-I$ residuals.

\section[]{Results for variables}

\begin{figure}
\begin{center}
   \leavevmode
   \epsfxsize=84mm
   \epsfbox[71 54 691 516]{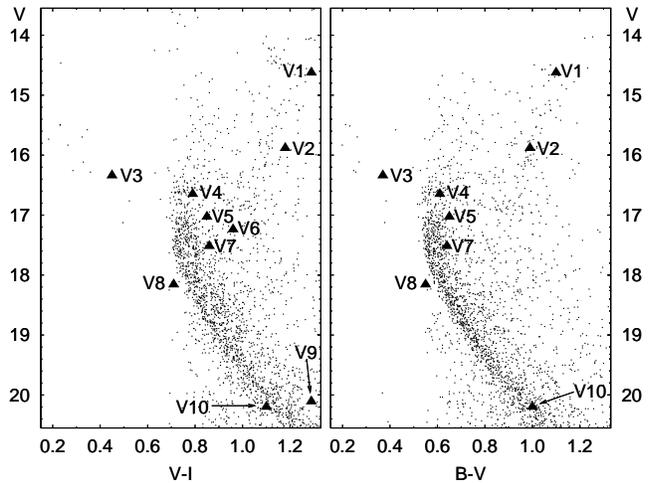}
\end{center}
\caption{Color-magnitude diagrams of Melotte 66 field with marked positions of variables.}
\label{stars_at_cmd}
\end{figure}

\begin{figure*}
\begin{center}
   \leavevmode
   \epsfxsize=16cm
   \epsfbox[136 240 505 630]{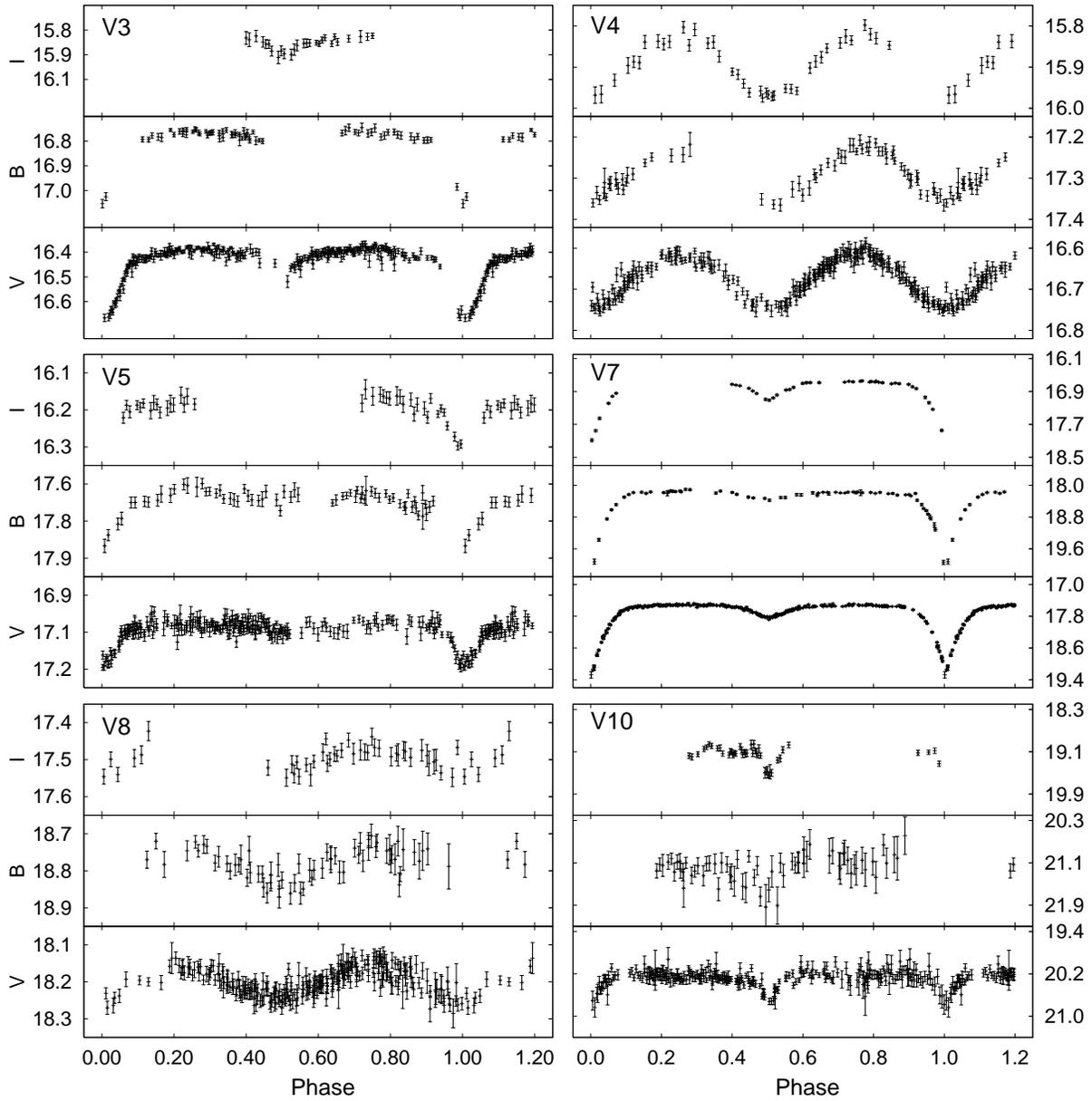}
\end{center}
\caption{$BVI$ phased light curves of periodic variables V3, V4, V5, V7, V8 and V10.}
\label{var1}
\end{figure*}

\begin{figure*}
\begin{center}
   \leavevmode
   \epsfxsize=16cm
   \epsfbox[136 387 505 482]{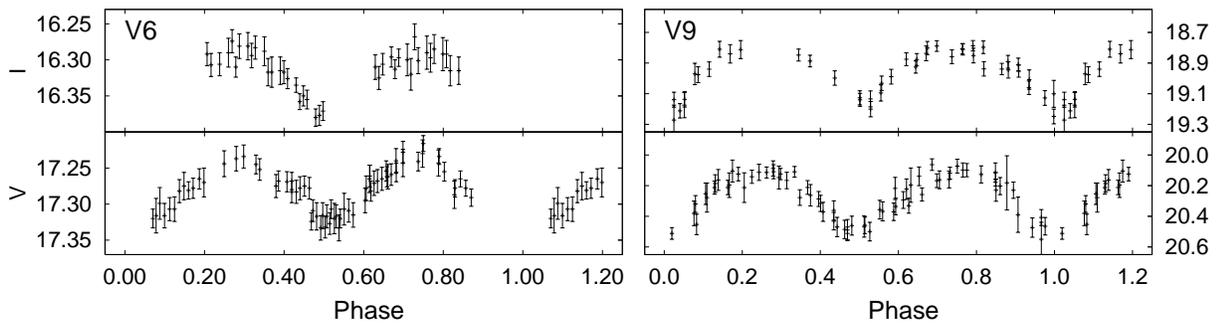}
\end{center}
\caption{Phased light curves of periodic variables V6 and V9.}
\label{var2}
\end{figure*}

\begin{figure*}
\begin{center}
   \leavevmode
   \epsfxsize=16cm
   \epsfbox[136 335 505 537]{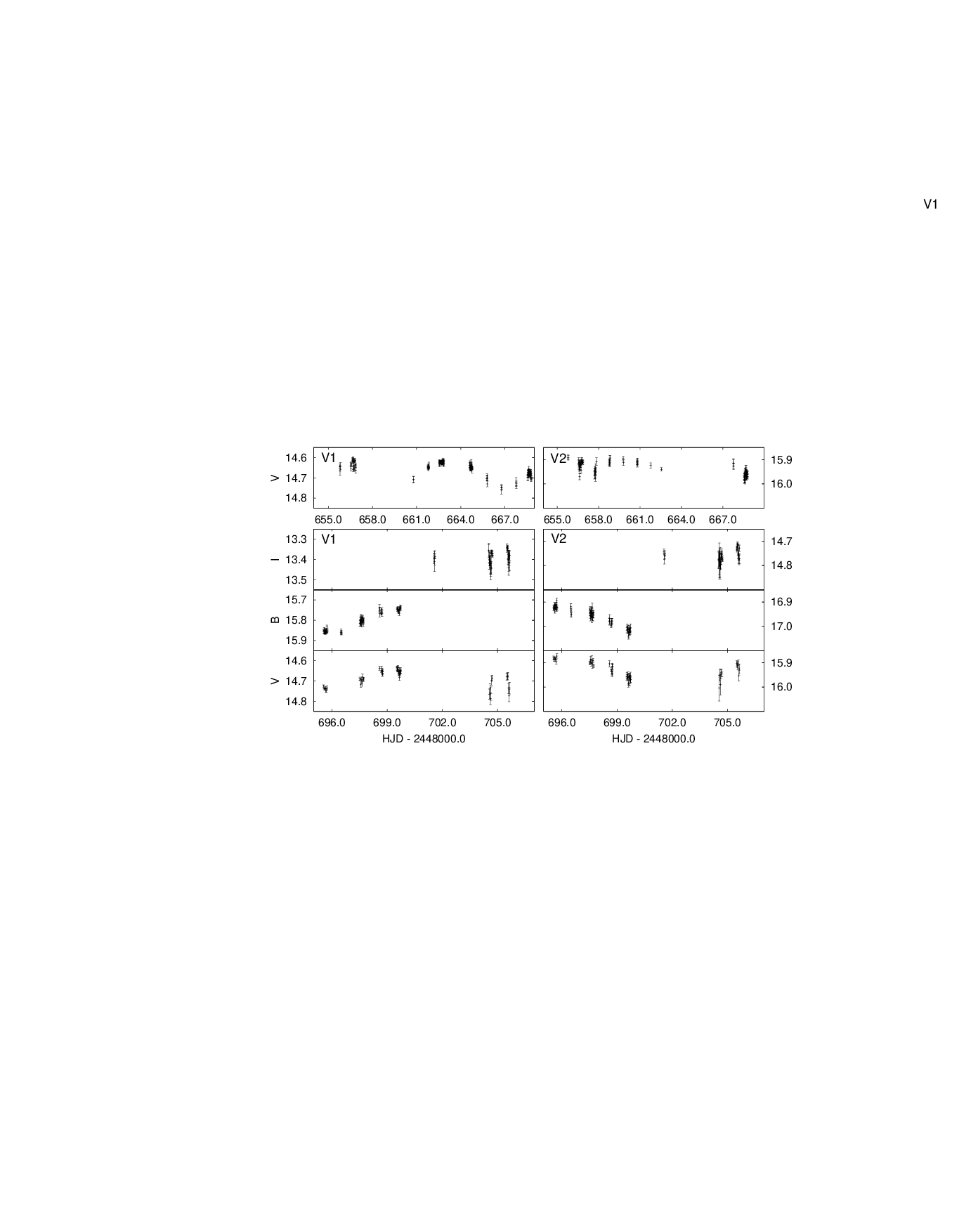}
\end{center}
\caption{Unphased light curves of variables V1 and V2.}
\label{var3}
\end{figure*}

Table 3 lists some basic characteristics of the light curves  of
variables V1--V10. For each star we list $V$ magnitude and colors
measured at maximum light. The full range of observed magnitudes in the
$V$ band is listed as $\Delta V$. The location of the variables on the
cluster $V/B-V$ and $V/V-I$ diagrams is shown in Fig.~\ref{stars_at_cmd}
while their light curves are presented in Figs.~\ref{var1}--\ref{var3}.
Stars V3--10 show periodic variability and can be classified as
eclipsing binaries (EC). Moreover, stars  V4, V6, V8 and V9 are likely
contact binaries (EW) based on their periods, colors and light curves.
Variables V1 and V2 are located on or near the cluster subgiant branch
while eclipsing binary V3 is a candidate blue straggler. Three other
non-contact binaries are located on or slightly above the cluster main
sequence. Unfortunately, none of these seem to be a good candidate for
spectroscopic follow-up aimed at a determination of cluster distance and
age. V5 shows shallow eclipses while V10 is too faint for high
resolution spectroscopy. The light curve of V7 is well defined and its
shape indicates that the system is  a likely semi-detached algol.

All variables are located inside the cluster radius which we estimate at
$9.\arcmin 2$ in Sec. 4. An examination of the angular distances listed
in Table 2 shows that most of the variables are in the central part of
Melotte 66.  There is radial velocity information available for only one
of the variables. Collier \& Reid (1987) list  $V_{\rmn r}$ = 75 $\pm$ 7
km s$^{-1}$ for V2=\#1210 based on a single observation.  The mean value
for 59 stars considered to be cluster members is $44\pm 12$~km s$^{-1}$.
This seems to indicate that V2 has a low probability to be a radial
velocity member of Melotte 66. However, this evidence is weak as the
radial velocity of V2 is likely to be variable.

Some indication of membership status can be provided for 4 contact
binaries. We have applied the absolute brightness calibration
established by Rucinski (2000) to estimate their absolute magnitudes.
The calibration gives $M_{\rmn V}$ as a function of unreddened color and
orbital period. We adopted  reddening of $E(V-I)=0.22$ resulting from
$E(B-V)=0.16$ advocated by Anthony-Twarog, Twarog \& Sheeran (1994). Using
dereddened  $V-I$ colors and periods listed in Table 3 we obtained
$M_{\rmn V}=3.51$, $M_{\rmn V}=3.07$, $M_{\rmn V}=3.79$ and $M_{\rmn
V}=6.33$ for V4, V6, V8 and V9, respectively. The apparent distance
moduli of these four variables follow from the observed values of
$V_{\rmn max}$. We obtained $(M-m)_{\rmn V}=13.14$, $(M-m)_{\rmn
V}=14.17$, $(M-m)_{\rmn V}=14.36$ and $(M-m)_{\rmn V}=13.78$ for V4, V6,
V8 and V9, respectively. These values  can be compared with the apparent
distance modulus of the cluster of $(M-m)_{\rmn V}=13.75$ as measured by
Kassis et al. (1997). The formal error of $M_{\rmn V}$ obtained from
Rucinski's calibration is about 0.3 mag and so the estimated distance
moduli are consistent with cluster membership for all four contact
binaries. This conclusion holds if we adopt $E(B-V)=0.23$ for the
cluster reddening as implied by maps of Schlegel et al. (1998).

\begin{table*}
\begin{minipage}{150mm}
\caption{Parameters of variables from the field of Melotte 66}
~
\begin{tabular}{@{}c l l l l l l l}
\hline
ID      &$V_{\rmn max}$	&$(V-I)_{\rmn max}$	& $(B-V)_{\rmn max}$	&$\Delta V$       		&$P$ [days]    	& $T_0$ HJD 2448000+	& Remarks  \\
\hline
V1      & 14.626        & 1.29                          & 1.10                          & 0.15:         & P $\sim$ 8$^d$ & --           & periodic?\\
V2      & 15.884        & 1.18                          & 0.99                          & 0.1:          & --            & --            & --            \\
V3      & 16.341        & 0.45                          & 0.37                          & 0.30          & 0.8015(2)     & 657.7128      & Ecl           \\
V4      & 16.65         & 0.79                          & 0.61                          & 0.17          & 0.4020(17)    & 656.6764      & Ecl-EW        \\
V5      & 17.03         & 0.85                          & 0.65                          & 0.15          & 0.7413(2)     & 658.7648      & Ecl           \\
V6      & 17.24         & 0.96                          & --                            & 0.12          & 0.6974(3)     & 657.6511      & Ecl-EW        \\
V7      & 17.519        & 0.86                          & 0.64                          & 1.78          & 0.5942(1)     & 656.5508      & {\small Ecl=V345~Pup}\\
V8      & 18.16         & 0.71                          & 0.55                          & 0.16          & 0.32903(6)    & 655.5672      & Ecl-EW         \\

V9      & 20.11         & 1.29                          & --                            & 0.49          & 0.2386(4)     & 656.7105      & Ecl-EW        \\
V10     & 20.2          & 1.1                           & 1.0                           & 0.9           & 0.8882(9)     & 668.7426      & Ecl           \\
\hline
\end{tabular}
\end{minipage}
\end{table*}

\subsection[]{Search for cataclysmic variables}

So far there are only 3 confirmed and one candidate cataclysmic
variables known in the whole sample of galactic open clusters: one in M
67 (Gilliland et al. 1991), two in NGC 6791 (Kaluzny et al. 1997) and
one in NGC 2158 (Mochejska et al. 2006). All of these clusters are old,
and as Melotte 66 is an old and rich cluster we decided to  search it
for possible cataclysmic variables. Two methods were used.

First we used the {\sc ISIS} image subtraction package (Alard \& Lupton
1998; Alard 2000) to look for objects showing outbursts. Cataclysmic
variables of dwarf novae type have average $M_{\rmn V}=7.5$ at
quiescence with outburst magnitudes spanning the range 2--8 mag (Warner
1995). At the cluster distance they would be observed at $V\approx 21$
at minimum light and at $13<V<19$  at maximum light. With a limiting
magnitude of  our observations of $V\approx 21$  it should be possible
to detect objects of this type while they are in outburst. Our search 
gave a negative result.

\begin{figure}
\begin{center}
   \leavevmode
   \epsfxsize=84mm
   \epsfbox[56 59 477 425]{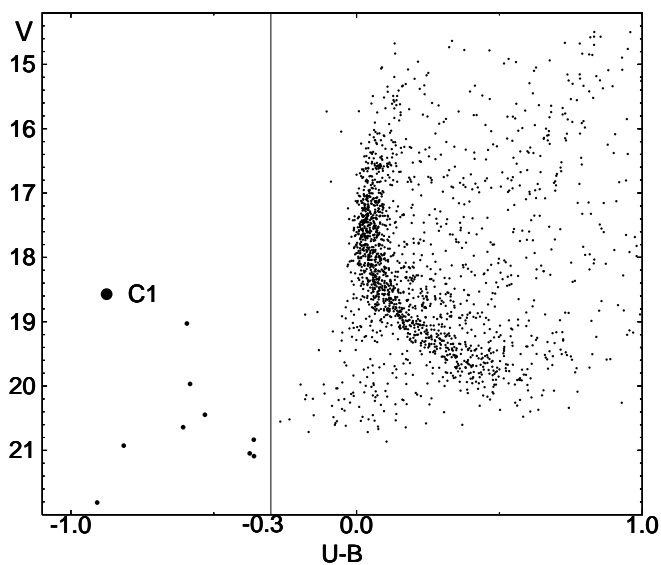}
\end{center}
\caption{$U-B$ versus $V$ color-magnitude diagram of Melotte 66.}
\label{subdwarf}
\end{figure}

The second method used relies on the fact that cataclysmic variables
have blue $U-B$ colors. In Fig.~\ref{subdwarf} we show a $V/U-B$ diagram
for the cluster field. It contains 10 objects with $U-B<-0.3$.  For 7 of
these objects we have time series photometry obtained with the FORD2
camera and for 2 photometry obtained with TEK2 camera. Examination of
these light curves showed than none of the objects shows any convincing
evidence for variability. The brightest of the blue objects, which we
denote C1, has $V=18.57$, $B-V=0.07$ and $U-B=-0.87$. The lack of
evidence for variability, including comparison of photometry from 1992
and 1999 seasons, indicates that  it is unlikely to be a quasar. It is
possible that the star is a hot subdwarf belonging to the cluster. 
We note that C1 is located very close to the projected cluster centre.
Very few confirmed hot subdwarfs are known in open clusters and it is of
interest to obtain a spectrum of C1 to clarify its nature. If C1 is a
member of the cluster then its absolute magnitude  is $M_{\rmn V}=5.4$.
This value corresponds to the faint end of the absolute magnitude
distribution observed for  hot subdwarfs in the field and in globular
clusters (Lisker et al. 2005; Moehler et al. 2002, 2004).

\section[]{Analysis of the CMD}

Some early photometric studies of Melotte 66 revealed an unexpectedly
large widths for the cluster subgiant branch and upper main sequence
(Hawarden 1976; Anthony-Twarog et al. 1979). Anthony-Twarog et al.
(1994) used $vbyH\beta$ CCD photometry to eliminate differential
reddening and variations in cluster metallicity as possible causes of
these large widths. They suggested a broad range of rotational
velocities among cluster stars as an explanation.

\begin{figure}
\begin{center}
   \leavevmode
   \epsfxsize=84mm
   \epsfbox[27 161 334 337]{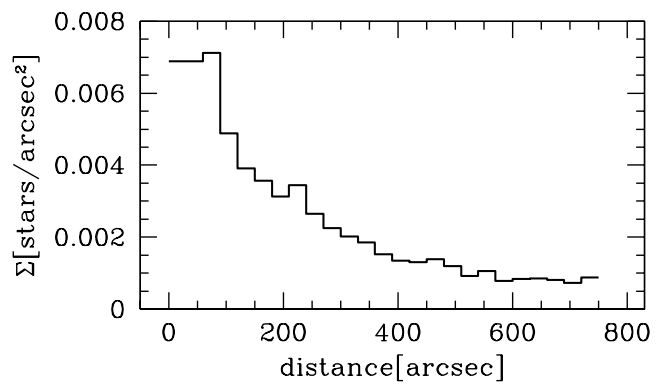}
\end{center}
\caption{Star density as a function of radial distance from the center of Melotte 66.}
\label{hist}
\end{figure}

\begin{figure}
\begin{center}
   \leavevmode
   \epsfxsize=84mm
   \epsfbox[62 65 500 388]{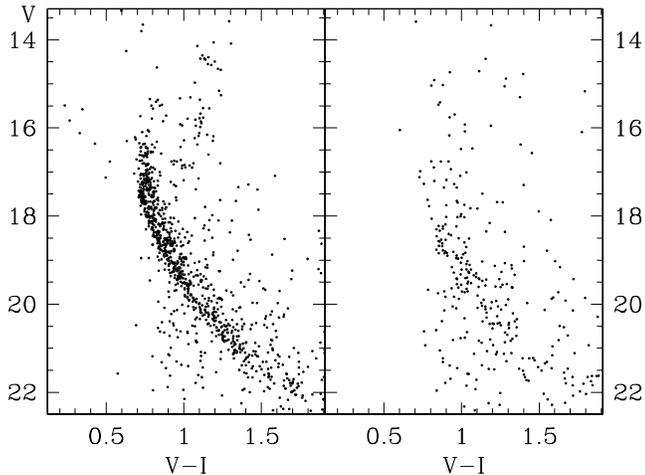}
\end{center}
\caption{The CMD's for the field covering central part of Melotte 66 (left) and for the "outer" field (right).}
\label{cmd}
\end{figure}

To further investigate this issue we started with a determination  of
the stellar density profile for the field observed with the SITE3 
camera. We observed 2134 stars with $16<V<21$ located near the cluster
main sequence in the $V/V-I$ CMD. The resulting density profile is shown
in Fig.~\ref{hist}. The projected density distribution flattens at a
radius of about 550 arcsec. The surface density of the field stars can
be estimated at $8.24\pm 0.11E-4$ stars per arcsec$^2$. This implies
that the SITE3 camera observations fail to cover the whole cluster E-W 
on the sky. By integrating  the density profile shown in Fig.~\ref{hist}
and subtracting the contribution from the field stars we conclude that
the cluster contains 1122 stars with $16<V<21$ or $2.8<M_{\rmn V}<7.8$ inside
a radius of $R=550$~arcsec. Note that this is a conservative lower limit
because no corrections for the incompleteness of the photometry have
been applied.  At a radius of  225 arcsec from the cluster centre the
cluster surface density is still a factor of 3 higher than the field
star density. In Fig.~\ref{cmd} we show $V/V-I$ CMDs for two groups of
stars: those lying inside a radius of $R=225$ arcsec from the cluster
centre and  those lying in the outer part of the observed field at
$R>550$ arcsec. The inner circle and the outer region cover equal areas
on the sky. For each star from the outer field a nearest match in the
CMD for the inner ring was located. Subsequently a  pair of stars
with the lowest separation was removed from both 
corresponding lists.
This procedure was continued  until it was impossible to
locate pairs with a separation $\delta V<0.25$ and $\delta(V-I)<0.15$.

\begin{figure}
\begin{center}
   \leavevmode
   \epsfxsize=84mm
   \epsfbox[52 65 474 388]{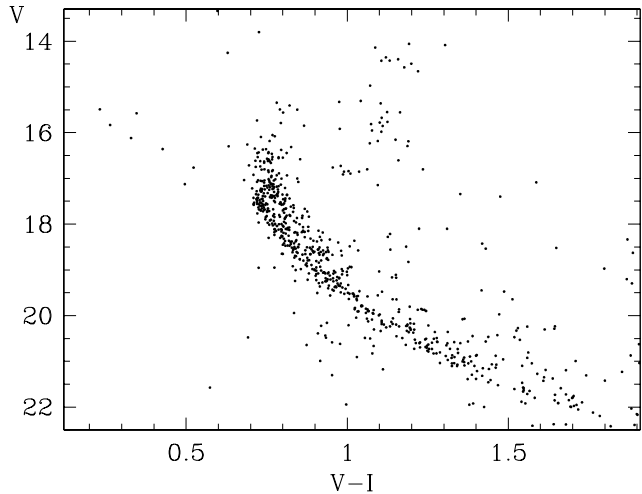}
\end{center}
\caption{Field-star corrected CMD for the central part of Melotte 66.}
\label{cmd_clean}
\end{figure}

The resulting "cleaned" CMD  for the inner region of Melotte 66 is shown
in Fig.~\ref{cmd_clean}. One may notice 7 candidate blue stragglers as
well as a clump of yellow stragglers at $V\approx 15.4$ and $V-I\approx
0.8$. The occurence of such objects in the cluster was first noted by
Hawarden (1976) and subsequently discussed by other investigators whose
papers are quoted above. Fig.~\ref{cmd_clean} shows that the cluster
main sequence is very sharply defined on the blue side. In particular,
for $17.4<V<19.5$ it is possible to distinguish a narrow, well defined
main sequence corresponding to a sample dominated by single stars. Above
this sequence is a second sequence consistent with a sample of binary
stars with mass ratios close to unity. It merges with the "single"
sequence near the turn-off region. As can be seen in
Fig.~\ref{stars_at_cmd}, the eclipsing binaries V4, V5, V7 and V9 are
located on the binary sequence.

The presence of a well populated binary sequence in the CMD of Melotte 66
has already been noticed by Kassis et al. (1997).

\begin{figure}
\begin{center}
   \leavevmode
   \epsfxsize=84mm
   \epsfbox[58 51 389 289]{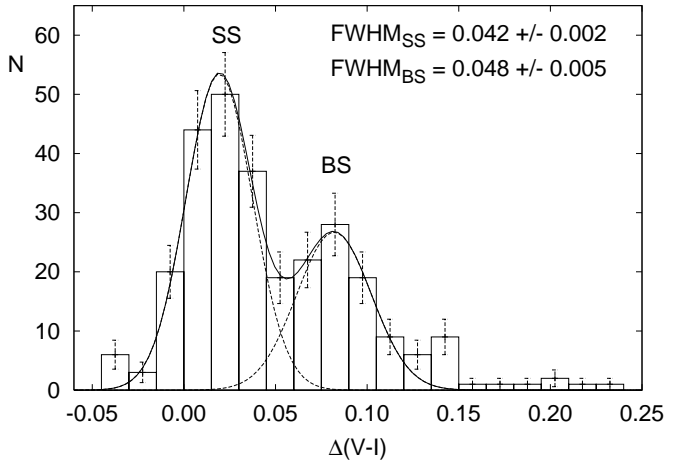}
\end{center}
\caption{Distribution of $\Delta (V-I)$ distances from the blue edge of
the cluster main sequence  (details in the text). SS and BS denote a
"single" and "binary" stars sequence respectively. Solid curve is a fit
to a double-gaussian. Broken curves correspond to distinct two gaussians
plotted with parameters derived in the double-gaussian fit.}
\label{ms_width}
\end{figure}

We have estimated the widths of these "single" and "binary" sequences
using stars from the cleaned CMD in a rectangle given by $17.5<V<19.5$
and $0.7<V-I<1.1$. The blue edge of the main sequence in this rectangle
was fitted by second order polynomial. For each star we have calculated
its distance (in color) from this polynomial. In Fig.~\ref{ms_width} we
show a histogram of star counts in  bins of 0.015 in $V-I$.
Fig.~\ref{ms_width} shows two peaks representing the two sequences. A
double-Gaussian fit to the two peaks indicates widths of $V-I$ = $0.042
\pm 0.002$ and $0.048 \pm 0.005$ for the "single" and "binary"
sequences, respectively. The two sequences are well resolved in over the
range $17.5<V<19.5$. At fainter magnitudes the sequences are smeared by
the uncertainties of the color measurements. At $16.0<V<17.5$ the
sequences cross each other leading to an apparent broadening of the
upper main sequence. Fig.~\ref{cmd_clean} shows that the cluster
subgiant branch is  well populated, with 4 or 5 stars scattered on the
blue side. We speculate that these stars represent an extension of the
binary sequence in the turn-off region.

\begin{figure}
\begin{center}
   \leavevmode
   \epsfxsize=84mm
   \epsfbox[18 64 349 354]{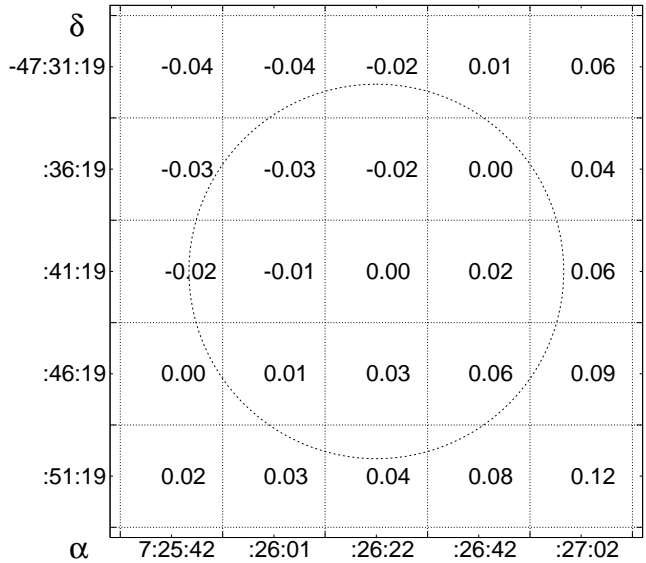}
\end{center}
\caption{Differential $E(B-V)$ reddening map with 5 arcmin resolution
in 25 $\times$ 25 arcmin box centred at the Melotte 66 coordinates.
Circle denotes cluster size R = 550 arcsec.}
\label{diff_ebv}
\end{figure}

We have checked how the $E(B-V)$ extinction varies across the field of 
the cluster  based on the Schlegel et al. (1998) reddening maps. In
Fig.~\ref{diff_ebv}. we show an $E(B-V)$ differential reddening map with
5 arcmin resolution in a 25 $\times$ 25 arcmin box centred on the
cluster. It is clear that even inside the cluster radius $\Delta E(B-V)$
can be as large as 0.09. According to Kassis et al. (1997) Melotte 66 is
located about $1.1$ kpc below the Galactic disk and 4.4 kpc from the
Sun. The line of sight to the cluster does not cross any outer spiral
arm of the Milky Way and so the observed gradients of $E(B-V)$ occur in
the interstellar matter located between the Sun and the
cluster.\footnote{ A possible extension of the Perseus spiral arm
(Caswell \& Haynes 1987) is present at the Galactic longitude of the
cluster at a distance of about 5~kpc. However, it is unlikely that this
contributes to the reddening in the direction of Melotte 66 due to the
large $z$ of the cluster.} Our data indicate the large width of the
cluster main-sequence can be explained by a high relative frequency of
binaries among the cluster stars together with differential reddening
across the cluster field.

\section{Conclusions}

We have detected a total of 10 photometric variables  in the field of
the old open cluster Melotte 66. Four out of eight eclipsing binaries
are contact systems which are probable members of the cluster. Like
other old open clusters Melotte 66 seems to have a rather high relative
frequency of contact binaries. Our estimate of the total number of
cluster members with $16<V<21$ at 1122 stars leads to a relative frequency of
EW stars of 0.36$\pm$0.18\%. Note, however, that the sensitivity 
of our survey for variables is rather low for $V>20$. 
This relative frequency can be
compared to that observed for field stars in the solar vicinity which is
estimated at 0.2\% for $3.5<M_{\rmn V}<5.5$ and at 0.1\% for $M_{\rmn
V}=6.0$ (Rucinski 2006). Three other binaries, including a candidate
blue straggler, are detached systems and one is probably a semi-detached
binary.  We also identified a promising candidate for a hot subdwarf
cluster member.

The projected stellar density profile was obtained for the cluster
field. We show that the angular radius of Melotte 66 reaches at least
550 arcsec. When corrected for the
contamination by field stars the cluster CMD shows a well defined and
narrow main sequence accompanied by a rich sequence of binary stars.

\section*{Acknowledgments}

This paper was supported by the  grant 1~P03D~001~28 from the Ministry
of Science and Higher Education, Poland. IBT acknowledges support from
NSF grant AST-0507325. AO acknowledges support from the Ministry of
Science and Higher Education, Poland grant no 1~P03D~006~27.



\end{document}